\documentclass[a4paper,12pt]{article}
\usepackage{graphicx}
\usepackage{float}

\newcommand{\be}{\begin{equation}}
\newcommand{\ee}{\end{equation}}   
\newcommand{\bea}{\begin{eqnarray}}   
\newcommand{\eea}{\end{eqnarray}}   
\newcommand{\non}{\nonumber}

\newcommand{\nn}{\nonumber\\}

\begin{document}

\title{Isospin equilibration in relativistic heavy-ion collisions
\protect\footnote{Supported by BMBF and GSI Darmstadt}\hspace{3mm}
\footnote{part of the PhD thesis of A. Hombach}}
\author{A. Hombach, W. Cassing and U. Mosel \\
        Institut f\"{u}r Theoretische Physik, Universit\"{a}t Giessen \\
        D-35392 Giessen, Germany}
\maketitle

\begin{abstract}

We study the mixing and the
kinetic equilibration of projectile and target nucleons 
in relativistic heavy-ion collisions
in the energy regime between 150~AMeV and 2~AGeV
in a coupled-channel BUU (CBUU) approach. 
We find that equilibrium in the
projectile-target degrees of freedom is in general
not reached even for large systems at low energy where elastic
nucleon-nucleon collisions dominate. Inelastic nucleon excitations
are more favorable for equilibration and their relative abundance
increases both with energy and mass.
Experimentally, the projectile/target admixture can be
determined by measuring the degree of isospin equilibration
in isospin asymmetric nuclear collisions.
For one of the most promising systems currently under
investigation, ${}^{96}_{44}{\rm Ru}+{}^{96}_{40}{\rm Zr}$, we investigate 
the influence of the equation of state and the inelastic in-medium 
cross section.

\end{abstract}
\vspace{2cm}
\newpage


\begin{section}{Introduction}
The aim of relativistic nucleus-nucleus collisions is to produce hot
and dense nuclear matter for sufficiently large space-time volumes
and to probe the properties of hadrons in a different environment
than the vacuum \cite{review1,review2,aich91,bass,review4,Prep}. 
From particle distributions, spectra and flow one can infer, e. g.,
on in-medium cross sections and the nuclear equation-of-state (EOS).
Experimentally,
the properties of the hot and dense part of the system are conventionally
described by a temperature $T$, baryon chemical potential $\mu_B$,
and expansion velocity $\beta$
characterizing the system at freezeout via the relative abundancy or
ratio of hadrons \cite{BM,reis97b}. However, if this thermal and chemical
equilibrium is actually achieved in the systems studied experimentally
is quite a matter of debate 
\cite{bass,reis97b,yvonne98,reis97,reis98}
and requires an analysis in terms of
nonequilibrium transport theory that involves the relevant hadronic
degrees of freedom and is applicable in a wide domain of bombarding
energies. Experimentally, such equilibration phenomena can be investigated
via the isospin degree of degree by using various systems with different
$N/Z$ ratios. This technique has been used especially at low bombarding
energies ($\leq 50$~AMeV \cite{MSU}) and is now being used also 
at higher energies up to the 1.5 A GeV range at the SIS \cite{yvonne98}. For a
recent review on the low energy studies we refer the reader 
to Ref. \cite{Wbauer}.

In this work we will investigate systematically the degree of 
isospin equilibration in nucleus-nucleus collisions from 
150~AMeV to 2~AGeV.
The time evolution of the systems we describe 
within the coupled-channel BUU (CBUU) transport approach \cite{teis96}
which is briefly described in Section 2. 
In Section 3 we analyse the dependence on the system mass and 
beam energy as well as on the centrality of the reaction and compare to
related simulations for equilibration phenomena in a finite box with
periodic boundary conditions, which allows to determine equilibration
times explicitly.
Detailed predictions for the total charge ratio for the system 
${}^{96}_{44}{\rm Ru}+{}^{96}_{40}{\rm Zr}$
at 400~AMeV and 1.5~AGeV -- which are presently being studied 
experimentally at the SIS -- follow in Section 4 while 
a summary and a discussion of open problems is given in Section 5.

\end{section}


\begin{section}{The CBUU-Model}
For our present study we employ the CBUU-transport-model \cite{teis96} to 
describe the time evolution of relativistic heavy-ion collisions. 
In this approach, apart from the nucleon, all
nucleon resonances up to masses of 1.95~GeV/c$^2$ are taken 
into account as well as the mesons $\pi$, $\eta$, $\rho$ and $\sigma$,
where the $\sigma$-meson is introduced to describe correlated 
pion-pairs with total spin $J = 0$. For the 
baryons as well as for the mesons all isospin degrees of freedom are 
treated explicitly. The hadrons included in our model obey a set 
of coupled transport equations for their one-body phase-space distributions  
$f_i(\vec{r},\vec{p},t)$ \cite{bertsch88,cassing88,cassing90,kweber93}:
\bea 
\label{buueq}
& & \frac{\partial f_1}{\partial t} + 
\left\{ \frac{\vec{p_1}}{E_1} + 
\vec{\nabla}_p\, U_1(\vec{r},\, 
\vec{p_1}) \right\} 
\, \vec{\nabla}_r f_1 
- 
\vec{\nabla}_r U_1(\vec{r},\, \vec{p_1}) 
\vec{\nabla}_p f_1 \nn
& & \hspace{1cm}  =  \sum_{2,3,4}
\frac{g}{(2\pi)^3} \, 
\int\! d^3 p_2 \, \int\! d^3p_3 \, \int\! d \Omega_4 \,\,\, 
\delta^3 \left(\vec{p}_1 + \vec{p}_2 
- \vec{p}_3 - \vec{p}_4 \right) \nn
& &\hspace{1.4cm} \times\,\,(
v_{34}\, \frac{d\sigma_{34 \rightarrow 12}}{d\Omega}\, 
f_3\, f_4\, \bar{f}_1\,\bar{f}_2 -
v_{12}\, \frac{d\sigma_{12 \rightarrow 34}}{d\Omega}\, 
f_1\, f_2\, \bar{f}_3\,\bar{f}_4)\quad.\hspace{1.4cm}\mbox{}
\eea
The l.h.s. of eq. (\ref{buueq}) represents the relativistic 
Vlasov-equation for hadrons moving in a 
momentum-dependent field $U_i(\vec{r}, \vec{p_i})$ \cite{teis96}, 
where $\vec{r}$ and $\vec{p_i}$ stand for the spatial and momentum 
coordinates of the hadrons, respectively. 
Note, that all space arguments are equal since the collision term 
in (\ref{buueq}) is local.
In our model the mesons are propagated as free particles, i.e.
$U_i(\vec{r},\vec{p_i}) \equiv 0$; this has been shown to be a valid
approximation at least for $\pi$ and $\eta$ mesons \cite{brat98}
whereas the in-medium properties of the $\rho$-meson are heavily debated
\cite{Prep}.

The r.h.s. of eq. (\ref{buueq}) describes the  
changes of $f_i(\vec{r}, \vec{p_i}, t)$ due to two-body collisions 
among the hadrons 
and two-body decays of baryonic and mesonic resonances. 
The in-medium collision rate is represented by 
$v_{12}\frac{d\sigma_{12 \rightarrow 34}}{d\Omega}$, 
where $\sigma$ denotes the free cross section 
and $v_{12}$ is the relative velocity between 
the colliding hadrons in their center-of-mass system.
The cross sections for the different channels are chosen to
reproduce the NN elementary $1\pi$-, $2\pi$- , $\eta$ and
$\rho$-production cross sections.
Taking elastic collisions only, 
$v_{12}\frac{d\sigma_{12 \rightarrow 34}}{d\Omega}$ equals
$v_{34}\frac{d\sigma_{34 \rightarrow 12}}{d\Omega}$;
in case of inelastic collisions 
we factorize the $N+N \to N+{\rm Resonance}$ cross section into 
(matrix element $\times$ phase space factor) and
use these matrix elements for the determination of the 
backward reaction \cite{teis_dr} thus avoiding the usual
detailed balance prescriptions.
For the most important nucleonic resonance, the $\Delta(1232)$,
we use a parametrization in line with the result of an OBE model
calculation by Dimitriev and Sushkov \cite{dimi86}.

In eq. (\ref{buueq}), $\bar{f}_i = 1 - f_i\, (i = 1, .., 4)$ 
are the Pauli-blocking factors for fermions. In the collision integrals 
describing two-body decays of resonances one has to replace the product 
(relative velocity $\times$  cross-section $\times \ f_2$) 
by the corresponding decay rate and introduce the proper fermion blocking
factors in the final channel. The factor $g$ in 
eq.~(\ref{buueq}) stands for the 
spin degeneracy of the particles participating in the collision 
whereas $\sum_{2,3,4}$ stands for the sum over
the isospin degrees of freedom of particles 2, 3  and 4.

We include the following elastic and inelastic 
baryon-baryon, meson-baryon collisions and meson-meson collisions: 
\bea 
N N & \longleftrightarrow & N N \nonumber \\
N N & \longleftrightarrow & N R \non \\
N R & \longleftrightarrow & N R'\non \\
N N & \longleftrightarrow & \Delta(1232) \Delta(1232) \non \\
R & \longleftrightarrow & N \pi \non \\
R & \longleftrightarrow & N \pi \pi \non \\
  & \quad = & \Delta(1232) \pi,\, N(1440)\pi, \, N \rho,\, N \sigma \non \\
N(1535) & \longleftrightarrow & N \eta \non \\
N N & \longleftrightarrow & N N \pi \non  \\
\rho & \longleftrightarrow & \pi \pi \, \,\, \, \,  \mbox{(p-wave)}\non \\
\sigma & \longleftrightarrow & \pi \pi \, \, \, \, \, \mbox{(s-wave)}, \label{reacs} 
\eea
where $R$ and $R^\prime$ represent the baryonic resonances \cite{teis96}. 
For the $NN\to\Delta\Delta$ reaction
we use the parametrization of Huber and Aichelin \cite{aichelin94};
$NN\to NN \pi$ is an s-wave contribution to the total pion 
production cross section below the Delta resonance.

The mean field $U$ entering the l.h.s. of eq. (\ref{buueq}) 
is of the MDYI-type as proposed by Welke et al.
\cite{welke88,gale90,zhang94}
\be
U(\vec{r}, \, \vec{p}) = 
A \frac{\rho(\vec r)}{\rho_0} + B \frac{\rho(\vec r)}{\rho_0}^\tau + 
2 \frac{C}{\rho_0} \int d^3 p'\frac{f(\vec{r},\vec{p'})}
{1 + \left(\frac{\vec{p}-\vec{p'}}{\Lambda}\right)^2}.
\label{welkepot}
\ee
This ansatz enables to guarantee energy conservation
since it can be derived from a potential energy density functional.
The parameters of the potential $U$ are chosen to match the 
requirements
\bea
&&\left.\frac{E}{A}\right|_{\rho_0}\!\! = {\rm -16\;MeV},\quad
\left.\frac{\partial E}{\partial \rho}\right|_{\rho_0}\!\! = 0, \quad
K = 210|260|380\;{\rm MeV},\quad \non\\[10pt]
&&U(E=300\;{\rm MeV})=0\;,\;\mbox{and}\quad U(p=\infty)=+32\;{\rm MeV} \;.
\label{potprop}
\eea
These constraints we derive from the results of the microscopic 
calculations by Wiringa et al. \cite{wir88_1,wir88_2}
with Hamiltonians required to describe NN scattering data, 
few body binding energies and nuclear matter saturation properties.
Best agreement to \cite{wir88_2} over the whole density regime 
of 0.1 to 0.5~fm$^{-3}$ would be achieved using a compressibility 
$K\simeq 230$ MeV, however, we fit the three different 
compressibilities denoted in (\ref{potprop}) as soft, medium and hard 
equation of state (EOS).
Additionally, we introduce an asymmetry potential 
$U_{\rm sym}$, depending only on the densities of protons and neutrons
\be
U_{\rm sym} = D \,\, \frac{\rho_p - \rho_n}{\rho_0} \,\tau_z\quad,
\ee
with D = 30~MeV and $\tau_z = \pm 1$ for protons and neutrons, 
respectively. With the help of this potential we are able
to reproduce the isospin and mass dependence of
the binding energy of nuclei correctly \cite{weid97}.
However, we find the effects of different binding energy and 
asymmetry potential to be negligible for the energy range 
($E/A \geq 400$ AMeV) considered here.

The baryons and mesons are represented by testparticles and
eq. (\ref{buueq}) is solved numerically in the parallel ensemble
algorithm. The particles are propagated according to
\bea 
\frac{d \vec{r}_i(t)}{d t} & = & \frac{\vec{p_i}}{E_i} + 
\vec{\nabla}_p \; U(\vec{r}_i,\, \vec{p}_i(t) ) \non \\
\frac{d \vec{p}_i(t)}{d t} & = & - 
\vec{\nabla}_r \; U(\vec{r}_i,\, \vec{p}_i(t) ) - q_i \vec{\nabla} 
V_C(\vec{r_i}), \label{tpeoms}
\eea
whereas for the collisions the Kodama-algorithm is used
requiring $b<b_{max}=\sqrt{\sigma/\pi}$
in addition to $b$=minimal in the current timestep
\cite{kodama}.

This model has been shown to adequately describe pion spectra
\cite{teis96}, pion multiplicities \cite{teis_dr}
as well as Coulomb effects on charged pion spectra \cite{teis_coulomb}.
\end{section}
%

\begin{figure}[!ht]
\centerline{\includegraphics[angle=90,width=18cm]{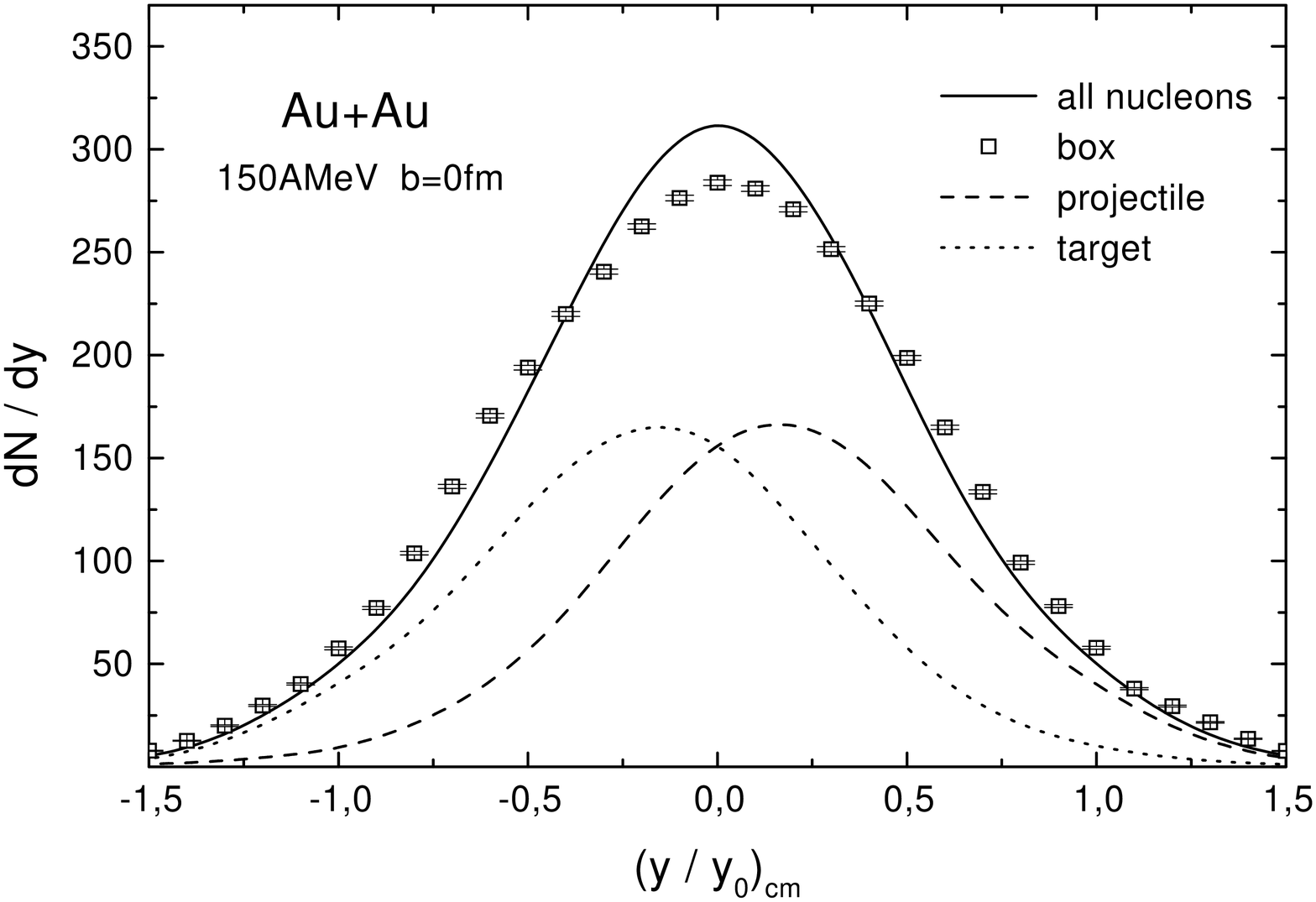}}
\caption{}
\label{dndy_150}
\end{figure}

\begin{figure}[!ht]
\centerline{\includegraphics[angle=90,width=18cm]{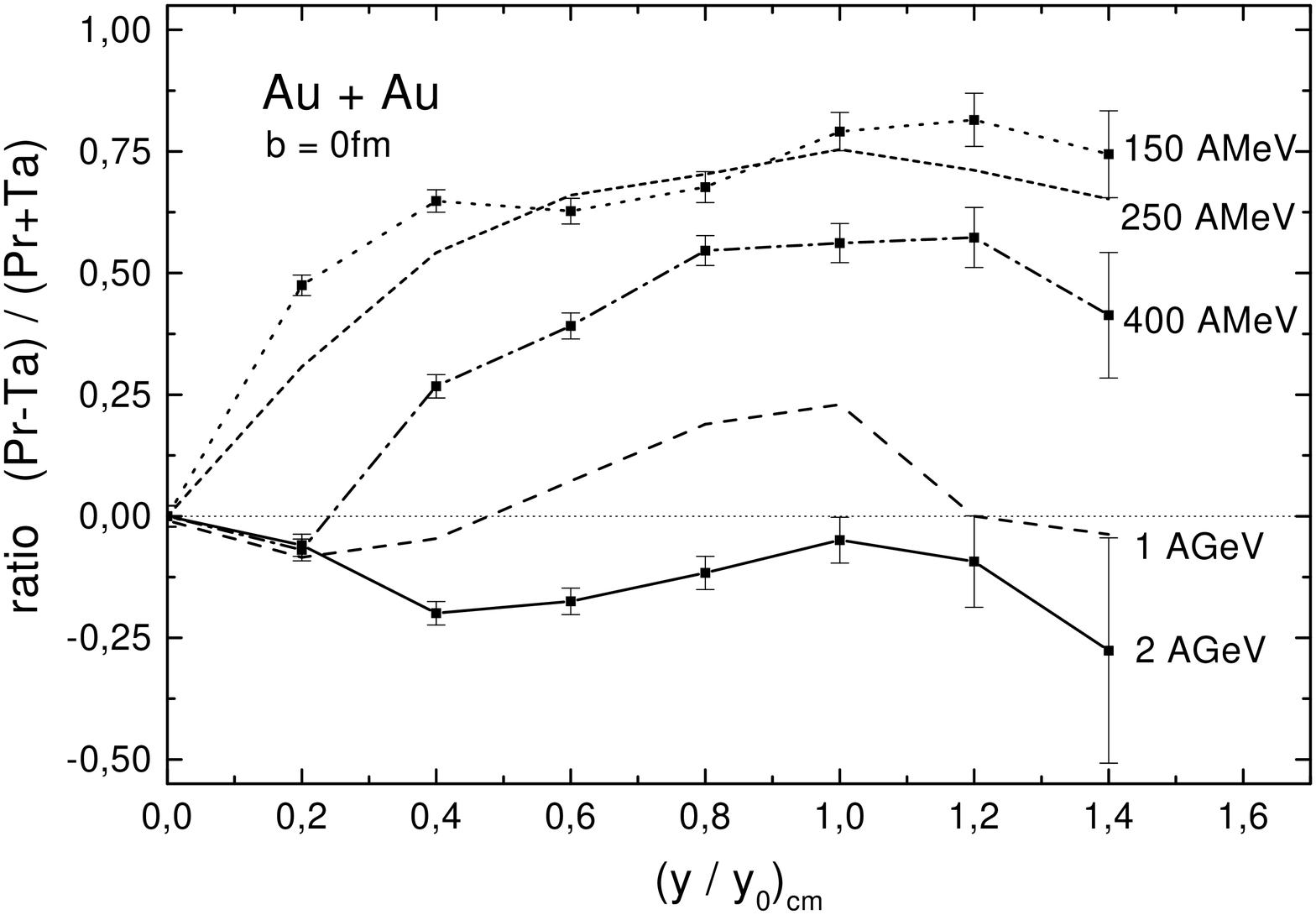}}
\caption{}
\label{transp_e}
\end{figure}

\begin{figure}[!ht]
\centerline{\includegraphics[angle=90,width=18cm]{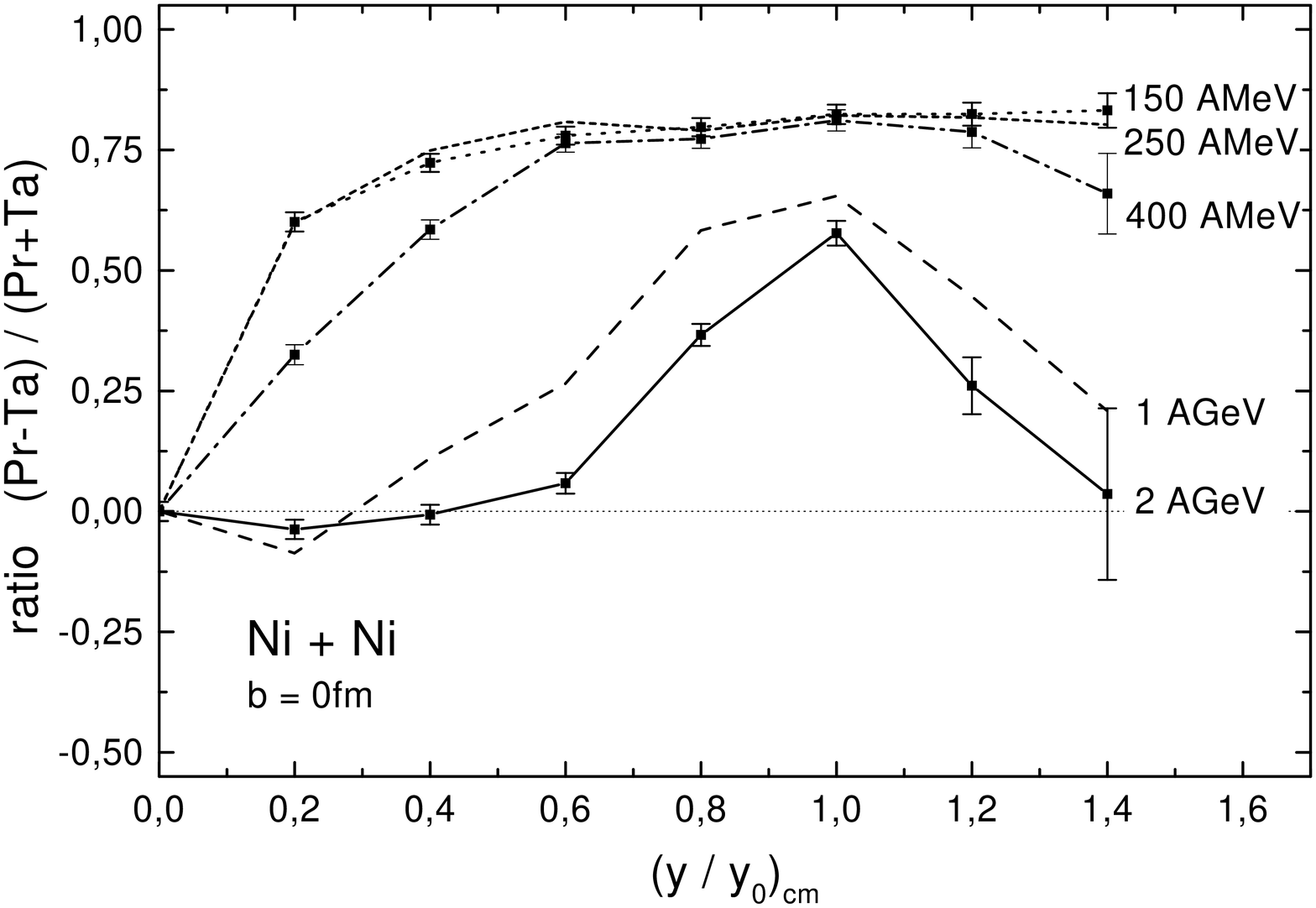}}
\caption{}
\label{transp_e_ni}
\end{figure}

\begin{section}{Equilibration in  heavy-ion collisions}

Within the CBUU model outlined in the previous section we first
investigate the isospin equilibration as a function of the
system size and incident energy. Later we will focus on 
the difference between the 'true' equilibration 
around the symmetry axis of the system 
and the experimentally accessible 
signal which is affected also by
surface effects and reaction centrality.
Note, that the term 'equilibration' in this work does not imply
thermal or chemical equilibrium;
equilibrium here means the total isotropy of the baryonic isospin
distribution. This charge isotropy is thought to be
synonymous to a total mixture of target and projectile
nucleons. 

We have chosen ${}^{58}_{28}{\rm Ni}+{}^{58}_{28}{\rm Ni}$ 
(with a total mass number of 116)  as a 
representative for a lighter system and 
${}^{197}_{\;79}{\rm Au}+{}^{197}_{\;79}{\rm Au}$ 
(with a total mass number of 394) as a heavy system. 
The experimentally investigated combination 
${}^{96}_{44}{\rm Ru}+{}^{96}_{40}{\rm Zr}$ \cite{yvonne98}
which we refer to with its total mass number of 192 is inbetween 
these two and will be considered in section \ref{ruzr}.
As the combinations Ni+Ni and Au+Au 
consist of reaction partners of equal isospin,
we will directly use the normalized ratio of target over projectile
nucleons
as a measure for the isospin mixture of the system. Note, that in the
transport approach each hadron trajectory can be traced back to the
incident projectile and target. Isospin equilibration or system admixture
can be analysed in the full phase space at every time. 

For relative comparison we will also
perform calculations in a finite box with periodic boundary conditions
using two shifted Fermi spheres with given $N/Z$ ratio as initial condition.
The relative momentum shift and $N/Z$ ratio is uniquely determined
by the projectile-target combination and bombarding energy, respectively.
These "infinite nuclear matter" studies allow to investigate the time scales
for kinetic and chemical equilibration for times $t \rightarrow \infty$, 
whereas any heavy-ion reaction is limited by the total interaction time. 

\begin{subsection}{Mass and energy dependence}
\label{mass_energy}

Fig.~\ref{dndy_150} shows the rapidity distribution in the final state 
of a central Au+Au collision at 150~AMeV. Though the total rapidity spectrum 
$dN/dy$ is close to that of a thermal distribution \cite{reis97},
here compared to a calculation in a box with periodic boundary conditions 
and $\langle\rho\rangle \simeq 0.9\times\rho_0$ for $t \rightarrow \infty$
(for a discussion of the choice of the box-size see below), 
the rapidity distributions of 
target and projectile nucleons are still clearly separated in the
'free' collision (in the following we will use the expression 'free'
collision for realistic simulations of nucleus-nucleus collisions)
within our semiclassical phase-space simulation.
This separation of projectile and target nucleons will 
approximately persist in a fully quantum-mechanical treatment
of the reaction dynamics
since only about 1/4 of the NN collisions occur between pairs
of identical particles (same spin and isospin) which cannot
be distinguished in the final state due to the Pauli principle.
However, taken as an approximation for isospin equilibration
the distributions in Fig.~\ref{dndy_150} show clearly that full
target-projectile-mixture is not reached in this reaction. 
This "transparency" is also supported e.g. by the longitudinally elongated 
event topologies for $Z=3$ fragments as  shown in Ref. \cite{reis97}.

We now consider -- in order to exclude surface effects -- a tube around 
the beam axis of radius r=1 fm and calculate
for each rapidity bin of typically $0.2\times y_0$ (where $y_0$ denotes
the cms rapidity normalized to the projectile rapidity in this system)
the number of 
projectile minus target nucleons and normalize it to the total number of 
nucleons in this bin. The resulting ratio then varies between +1 for 
having only projectile nucleons in one bin and 
-1 for target nucleons, respectively, and is a direct measure for the 
degree of target/projectile mixing or isospin equilibration 
in the system.

This target/projectile ratio is shown in Fig.~\ref{transp_e} 
for central Au+Au collisions at different kinetic energies. 
As already indicated in Fig.~\ref{dndy_150}, the ratio is clearly different
from zero at low energies. Above about 400~AMeV the Au+Au system begins to 
equilibrate and shows even a repulsion of target and 
projectile matter at 2 AGeV. 
This increase of equilibration in the target/projectile ratio with 
energy, however, does not correspond to an increase of 
stopping of the system, measured by the quantity $E_{\rm rat}$,
which gives the transverse energy to longitudinal energy ratio 
\begin{equation}
\label{erat}
E_{\rm rat} = \frac{\sum_i \left(p_x^2 + p_y^2\right)}{\sum_i p_z^2}
\quad.
\end{equation}
The mean values of the $E_{\rm rat}$ distributions for the
Au+Au systems are shown in Table~\ref{erat_values} and indicate 
that the maximum stopping is already reached around 400~AMeV
incident energy. Thus the isotropy of the isospin distribution is not 
a trivial effect related to the kinetic deceleration of the nuclei 
throughout the collision.

\begin{table}[H]
\vspace*{1ex}
\begin{center}
\begin{tabular}{c|c|c}
$E_{\rm kin}$ [MeV] & $\qquad x_0\qquad$ & $\qquad\Delta x\qquad$ \\
\hline\hline
2000 & 1.22 & 0.33 \\
1000 & 1.50 & 0.36 \\
400  & 1.92 & 0.51 \\
250  & 1.74 & 0.50 \\
150  & 1.71 & 0.49
\end{tabular}
\end{center}
\caption{Mean values of the $E_{\rm rat}$ distributions from 
central Au+Au collisions, obtained from a gaussian fit with
the parameters $x_0$ and $\Delta x$ for offset and width,
respectively.}
\label{erat_values}
\end{table}

A very similar behaviour of the target/projectile ratio is found 
for the Ni+Ni system, shown in Fig.~\ref{transp_e_ni}, 
besides the fact that this lighter system 
never fully admixes. At projectile rapidity the number of
projectile nucleons always exceeds the number of target nucleons. 
This is due to the lower size of the fireball which limits the number of
inelastic collisions and is too small for an isospin equilibration of the
\mbox{Ni+Ni} system.

In order to investigate the influence of the size and temperature 
of the fireball on isospin
equilibration in more detail we have calculated the equilibration ratio
as function of time for systems in a box with periodic boundary conditions.
The nuclei are initialized on top of each other in coordinate space and given
relative momenta according to the energy of the reaction.
We compare these calculations to the time available for equilibration 
during the lifetime of the fireball in a 'free' collision. 
The box was chosen such that the average density  lies between
0.9 and $0.7 \times \rho_0$ for Au+Au and Ni+Ni, respectively.
This density was chosen since it corresponds to the 
central density at the hadronic freeze-out time \cite{hom98_2}
and the size of the box is large enough to expect the evolution of the 
collision in the initial phase to be roughly the same as without periodic
boundaries.
A change of the box-size would, of course, lead to different
internal quantities like temperature and collision rate 
or measurable quantities like particle spectra and rapidity distributions.
Note, that in Fig.~\ref{dndy_150} the broader $dN/dy$-distribution
of the box calculation compared to the one of the 'free' collision is a hint 
to a higher final energy density of the box.
However, this is not crucial for our investigation here, since an
increase of the size of the box would only 
lead to a lower collision rate
in the late collision stage and thus even increase the time
necessary for equilibration.
As global equilibration ratio for this box calculations we have averaged 
the individual target/projectile ratios 
in each rapidity bin over all rapidities from $-0.9\, y_0$ to $0.9\, y_0$
times sgn$(y)$.
The 'free' fireball lifetimes to compare with
we define as the time interval from the rise of the density
in the center of the reaction zone above $\rho_0$ to the drop
below $0.5 \rho_0$ in a 'free' collision. 
This corresponds to the time period in which
practically all NN-collisions in a HIC happen as indicated by 
Fig.~\ref{au2dens-coll}.

The results for both Au+Au and Ni+Ni are shown in Fig.~\ref{eq_auau_nini}.
All global ratios start from zero
(since there are no particles in the rapidity interval $-0.9<y<0.9$ initially)
and reach a maximum between 5 and 10~fm/c. The maximum possible value of 
the global ratio is 19. Defining equilibration by the ratio $\leq 10\%$
of the maximum, clearly the time needed to equilibrate is much 
longer than the lifetime of the fireball at 150 and 400~AMeV for both
systems. On the other hand, equilibration is reached for 
${}^{197}_{\;79}{\rm Au}+{}^{197}_{\;79}{\rm Au}$ 
at 1 AGeV but not for ${}^{58}_{28}{\rm Ni}+{}^{58}_{28}{\rm Ni}$ 
at this energy.
At 2 AGeV a negative ratio is observed between t=8 and 15~fm/c,
which is due to the
strong repulsion of the target and projectile matter in the dense 
reaction zone. Since matter can not escape to the side due to the boundary 
conditions imposed, a shock front runs 
from the collision zone in opposite direction to the instreaming matter
thus inverting their average direction of motion. Less pronounced 
this happens also in a 'free' reaction as reflected by the negative ratio 
at 2 AGeV in Figs.~\ref{transp_e} and \ref{transp_e_ni}.

In summarizing this section, we like to point out that 
the main limiting factor for achieving
projectile-target equilibration in the energy region of 150~AMeV
to 2~AGeV is the number of {\it inelastic} NN-collisions
per volume that occur throughout a HIC. This collision number
is again limited by the lifetime of the fireball,
given by the size of the system, and the incident energy of the collision. 
We find the decrease with bombarding energy to be stronger than the
increase of the overall reaction time when going to lower 
bombarding energies. On the other hand,
the relative influence of the nucleon potential
increases with decreasing energy. Here especially the momentum dependence 
gives some additional stopping for the relative motion between target and
projectile as compared to a pure cascade or a density dependent
(Skyrme) potential. However, even at 150~AMeV the collisions
play the dominant role for stopping.

\end{subsection}
\begin{subsection}{Surface effects and dependence on centrality}
\label{surf_centr}

In the previous section we have considered the 'true' mixture of
the system by taking 
a tube around the beam axis, thus excluding any surface effects.
Experimentally, however, only an overall equilibration is measurable
and one has to consider the influence of corona effects,
which means that the surfaces of the colliding nuclei pass each other.
Additionally, the results measured will be affected by dominant
contributions from non-central collisions where spectator matter
will spoil the signal of equilibration in the fireball. So the ability
to determine central events will be a crucial point 
experimentally. In the following we will therefore investigate both
the effects of surface contributions as well as criteria for
centrality selection on the signal for the Au+Au system, 
for which we could extensively check the agreement of our results
on multiplicity, transverse energy and directivity distributions 
to experimental data at various energies
\cite{reis97, wien93}.

Fig.~\ref{mixing_b} shows the target to projectile ratio defined
in the last section for central, b=1~fm and b=2~fm collisions in
comparison to the 'true' ratio along the beam axis,
determined in the cylinder geometry as explained in the previous
section.
The b=0 overall
ratio is, besides a difference at high rapidities, in very good
agreement to the true ratio; for b=1~fm the agreement is still acceptable.
Surface effects contribute mainly at $y>y_{pr}$ and are
negligible for rapidities below $0.9 y_0$.
However, the centrality selection is crucial since already the b=2~fm
results differ substantially from the 'true' signal.
Since higher impact parameters enter in an impact parameter integrated
measurement with higher statistical weight, we have checked on
experimental possibilities to suppress these.

A usual way to determine central events is given by selecting events
with a high transverse energy to longitudinal energy ratio $E_{\rm rat}$
as defined in Eq.~(\ref{erat}),
and applying additional cuts in directivity
\begin{equation}
\label{direc}
D = \frac{|\sum \vec p_t |}{\sum |\vec p_t |} \quad, y>y_{cm} \;,
\end{equation}
where the usual FOPI-Cuts of $7^0 \le \Theta_{\rm lab} \le 30^0$
have to be used.

Fig.~\ref{erat_b} shows the experimental $E_{\rm rat}$ distribution
for Au+Au at 400~AMeV in comparison to the CBUU calculations.
Additionally, the CBUU distributions
obtained for b=0.1, b=1 and b=2 runs are shown separately.
The distributions of the individual impact parameters are very broad
and especially the b=2~fm  impact parameter 
still contributes substantially at high $E_{\rm rat}$.
Thus for an event sample with $E_{\rm rat}$ $> 1.2$ we have applied additional
cuts in directivity, ranging from $D<0.02$ to $D<0.3$.
The relative contributions of runs at different impact parameters are
shown in Fig.~\ref{direc_erat}. 
It can clearly be seen that requiring a narrow cut in $D$ is an
excellent way to select central events, though contributions
from b=2 fm runs cannot be completely excluded.
The latter we cannot quantify exactly 
due to statistical uncertainties
caused by the low number of events surviving these strong cuts.

However, the systems should be taken as massive as possible since 
corona effects increase with decreasing mass. For the Au+Au system 
investigated in this subsection they are negligible. Beyond this,
when requiring sufficiently low directivity $D$, it should
be possible experimentally to exclude non-central events and to
measure a signal very close to the 'true' target$-$projectile
mixture as discussed in the previous section.

\end{subsection}
\end{section}
\begin{section}{The Ru+Zr system}
\label{ruzr}

Experimentally projectile and target nucleons are indistinguishable.
In order to gain information on the degree of
equilibration one can choose projectile $-$ target combinations of different
isospin and evaluate the uniformness of the isospin distribution
at different rapidities.
Experiments are currently under consideration at the SIS,
where measurements with ${}^{96}_{44}{\rm Ru}$ and  ${}^{96}_{40}{\rm Zr}$
at 400~AMeV and at 1.5~AGeV are being investigated.

The system Ru+Zr with a total mass number of 192 behaves
more similar to the Ni+Ni system than to Au+Au as shown in 
Fig.~\ref{transp_m}.
For this system we have also first determined the target over projectile
ratio since this quantity is, first of all, numerically much more stable and,
second, direct proportional to the isospin ratio.
For the two energies mentioned above, we have
investigated the factors influencing the mixing ratio and the
dependence on varying these.
The results are shown in Fig.~\ref{ruzr_b0}.
At 400~AMeV (l.h.s.) we find the incompressibility  $K$
of the equation of state (EOS) to be the dominant factor,
even though the dependence is not very strong.
Close to  the Ni+Ni system discussed in
section \ref{mass_energy}, in the Ru+Zr system the number of
inelastic collisions
is much too low at 400~AMeV for equilibration thus giving rise to a
target/projectile ratio up to 0.6 at projectile rapidities.
By increasing the compressibility one decreases the number of collisions
even more; this leads to an increase of the ratio and vice versa.
Note that when varying the inelastic cross section within reasonable
limits ($\pm 30\%$) negligible effects are found in this energy regime
which is essentially dominated by elastic processes.

At 1.5 AGeV the situation is different. Though an overall
equilibration is again not reached here due to the small number of collisions
limited by the size and lifetime of the fireball, the target to projectile
ratio is close to zero up to normalized rapidities $y_0 \simeq 0.6$ and 
reaches its maximum value of
0.4 at $y_{pr}$. The incompressibility of nuclear matter plays a
minor role at these energies, whereas  a variation of the inelastic cross
section changes the degree of equilibration by $\approx 10\%$.
So in principle it could be possible to determine
in-medium cross sections and the EOS separately by looking at the
isospin equilibration at different energies and system sizes.

For the charge or isospin ratio one has to consider the shift of the
total baryonic isospin due to charged pion production at higher energies. 
In order to normalize
the p/n-ratio vs. rapidity to $\pm 1$ for ${}^{96}_{44}{\rm Ru}$
and ${}^{96}_{40}{\rm Zr}$, respectively, one has not to take the initial
isospin ratios of $\sim$0.85 for Ru and $\sim$0.71 for Zr but the final
ones given by 
$\frac{44-{\scriptscriptstyle\sum}\pi^+}
{96-44-{\scriptscriptstyle\sum}\pi^-}$ and
$\frac{40-{\scriptscriptstyle\sum}\pi^+}
{96-40-{\scriptscriptstyle\sum}\pi^-}$ for Ru and Zr, being
approximately 0.90 and 0.81 in our calculation at 1.5 AGeV.
The statistical uncertainties of our calculation on the isospin ratio  --
given not only by the number of
particles in each rapidity bin but also by the statistical uncertainty
of the charged pion number -- are shown in Fig.~\ref{ruru}.
These fluctuations blur the difference between the 400~AMeV
and the 1.5 AGeV curve shown in Fig.~\ref{ruzr_pn},
which should closely follow the respective lines in Fig.~\ref{ruzr_b0}.
However, experimentally it should be possible to determine the charge
ratios with sufficient accuracy.

\end{section}

\begin{section}{Summary}

In this paper we have explored the possibility to determine the
degree of target/projectile equilibration via the measurement of
isospin ratios vs. rapidity in the final state of HIC's of
isospin asymmetric systems.
The analysis has been performed within the coupled-channel (CBUU)
transport approach which has proven in Refs. \cite{teis96,teis_coulomb} 
to adequately describe pion spectra as well as baryon flow in
the SIS energy regime \cite{Pradip}.
In our calculations the isospin dependence of both the inelastic 
and elastic scattering cross section is taken into account, 
as well as an isospin dependent mean-field potential.
However, in the energy range considered here the 
influence of the isospin asymmetric potential term is negligible 
as well as a separate consideration of Pauli blocking for 
protons and neutrons. 

We find that full equilibration is practically never achieved even in
central collisions of Au + Au. This is due to the fact that elastic
collisions at lower energy ($\leq$~400~AMeV) are not very effective
for baryon stopping due to forward peaked angular distributions. Inelastic
baryon excitations help very much for equilibration such that central
Au+Au collisions at 1 AGeV show an approximate equilibration. This is
due to the fact that the lifetime of the fireball is sufficiently long
as compared to an equilibration time evaluated for a related infinite
nuclear matter problem within the CBUU approach. For Ni+Ni we find
no full equilibration at all bombarding energies considered here.

We have shown explicitly the influence of surface effects and examined
criteria for
centrality selection on the equilibration signature. 
As result, it should be possible, when selecting central events 
carefully, to achieve experimentally a signal at least close to 
the 'true' equilibration of the system, i.e. the equilibrium
in that region where the centers of the colliding nuclei are located.

In addition,
it has been shown that the incompressibility $K$ of the EOS 
modifies the signal at lower energies somewhat, 
whereas  medium modifications of the
cross section dominantly influence the signal at high bombarding energy.
Thus in principle it might be possible to determine the EOS and the
medium modifications of the inelastic cross section independently by
$N/Z$ ratios vs. rapidity in very central collisions of different systems
at e.g. 400~AMeV and 1.5~AGeV. 
For a determination of the EOS light systems should be used
since the short lifetime of the fireball limits the number of inelastic
collisions. One promising system could be $^{48}$Ca+$^{50}$Cr as 
proposed in \cite{bass}.
On the other hand, the determination of changes in the in-medium cross section
does not require very heavy systems in the mass region of 
${}^{197}_{\;79}{\rm Au}+{}^{197}_{\;79}{\rm Au}$ . 
These equilibrate anyhow
in the energy regime of $\approx$ 1~AGeV and thus are insensitive
to minor changes ($\pm$ 30\%) of the inelastic cross section. 
In this sense the Ru+Zr system is promising in yielding results 
limiting both the variety of the compressibility of the EOS and
the in-medium cross sections.

\end{section}

\pagebreak
\pagestyle{empty}
\begin{section}*{Figure captions}

{\bf Fig. \ref{dndy_150}}:
Baryon rapidity distribution of a central 150 AMeV Au+Au
collision (solid line)
in comparison to the thermal equilibrium rapidity distribution as calculated
in a box with periodic boundary conditions (squares).
For the central collision the
rapidity distributions of projectile nucleons (dashed)
and target nucleons (dotted) are displayed separately.\\

\noindent
{\bf Fig. \ref{transp_e}}:
Ratio of target to projectile nucleons in a tube along the beam direction
as defined in section \ref{mass_energy} 
as a function of rapidity for central Au+Au collisions at
different bombarding energies. For 150, 400 and 2000 AMeV
error bars are given to indicate the statistical uncertainty
of the result.\\

\noindent
{\bf Fig. \ref{transp_e_ni}}:
Same as Fig. \ref{transp_e} for the Ni+Ni system.\\

\noindent
{\bf Fig. \ref{au2dens-coll}}:
Evolution of the density in the center of the reaction zone
(solid line) and the collision rate (dotted line)
in a central Au+Au collision at 2 AGeV.
The vertical dashed lines indicate the boundaries of
the time interval available for equilibration as given
by the onset of the collisions and the drop of the
central density below $0.5\rho_0$.\\

\noindent
{\bf Fig. \ref{eq_auau_nini}}:
Global equilibrium ratio as a function of time for systems evolving in a box
with periodic boundary conditions; Au+Au (l.h.s.) and
Ni+Ni (r.h.s.) in comparison to the fireball lifetimes
of the corresponding 'free' collision (vertical dotted lines).\\

\noindent
{\bf Fig. \ref{mixing_b}}:
Comparison between the ''true'' equilibrium in a tube along
the z-axis (excluding surface effects) (solid line)
and the overall equilibrium for b=0, 1 and 2~fm collisions
(dashed, dash-dot and dotted line, respectively) for Au+Au at 0.4 AGeV.\\

\noindent
{\bf Fig. \ref{erat_b}}:
The distribution in $E_{\rm rat}$ for Au+Au at 400 AMeV.
The experimental data (squares) have been taken from \protect\cite{reis97}
while the CBUU results are given by the dotted line.
Also shown are the individual distributions for
b=0.1, 1 and 2~fm collisions from the CBUU calculations
(solid, dashed and dash-dotted line).\\

\noindent
{\bf Fig. \ref{direc_erat}}:
Contributions to different directivity bins of an event sample
with $E_{\rm rat}>1.2$ for Au+Au at 400 AMeV.\\

\noindent
{\bf Fig. \ref{transp_m}}:
Mass dependence of the target-projectile mixture.
Shown are target to projectile ratios vs. rapidity
for Au+Au (dashed), Ru+Zr (solid) and Ni+Ni (dotted).\\

\noindent
{\bf Fig. \ref{ruzr_b0}}:
Target to projectile ratio for the Ru+Zr system for different
energies of 400 AMeV (l.h.s.) and 1500 AMeV (r.h.s.).
The upper plots show the influence of different equations
of state while in the lower plots the in-medium inelastic cross section
is varied by $\pm 30\%$.\\

\noindent
{\bf Fig. \ref{ruru}}:
Charge ratios vs. rapidity for Ru+Ru (solid lines)
and Zr+Zr (dashed lines) collisions at
400 AMeV (upper panel) and 1500 AMeV (lower panel).\\

\noindent
{\bf Fig. \ref{ruzr_pn}}:
Charge ratios vs. rapidity for Zr+Ru collisions at 400 AMeV
(solid line) and 1500 AMeV (dashed line).\\

\end{section}
\pagebreak

\topmargin-2cm

\begin{figure}[!ht]
\centerline{\includegraphics[angle=90,width=18cm]{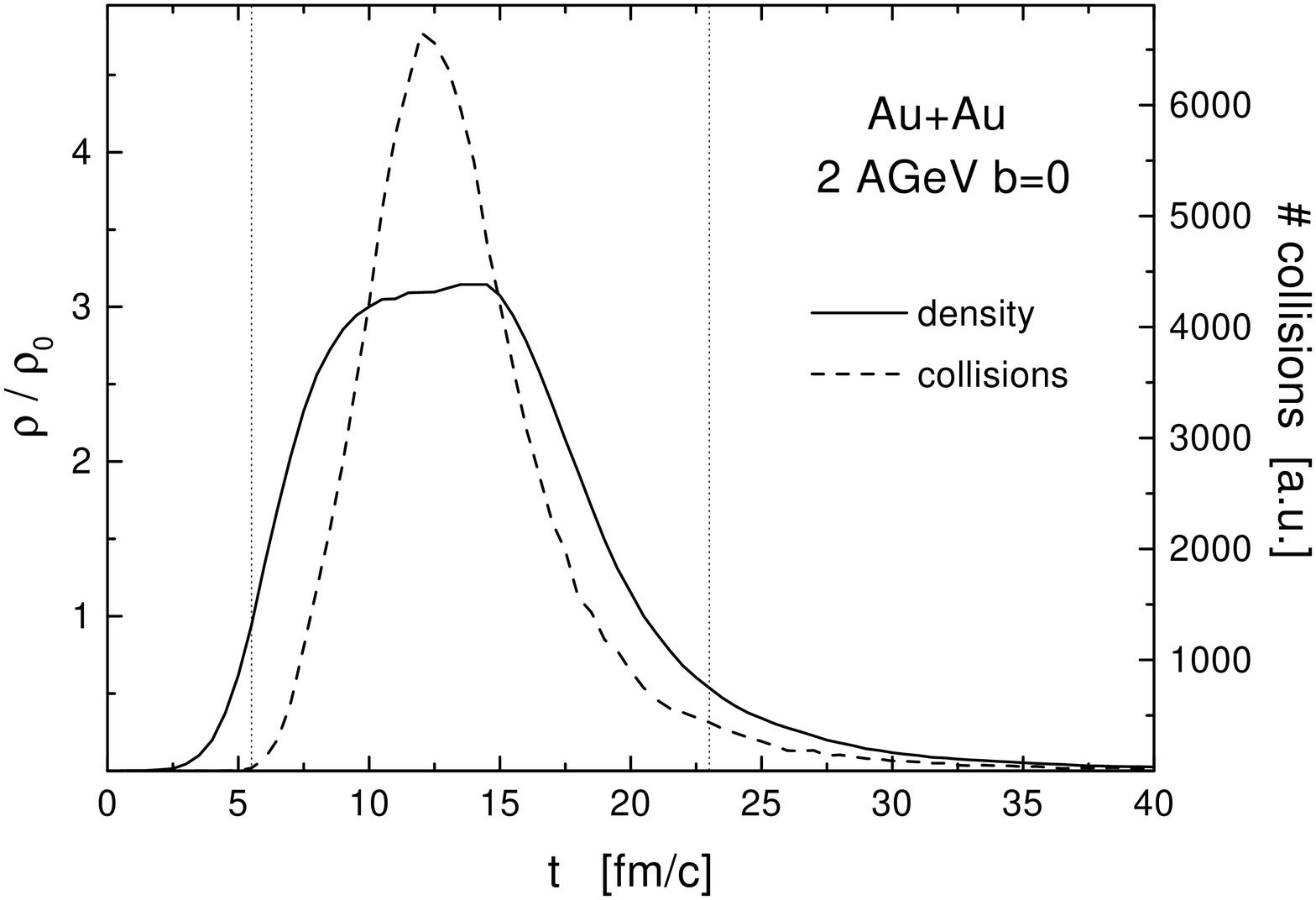}}
\caption{}
\label{au2dens-coll}
\end{figure}

\begin{figure}[!ht]
\includegraphics[angle=90,width=16cm]{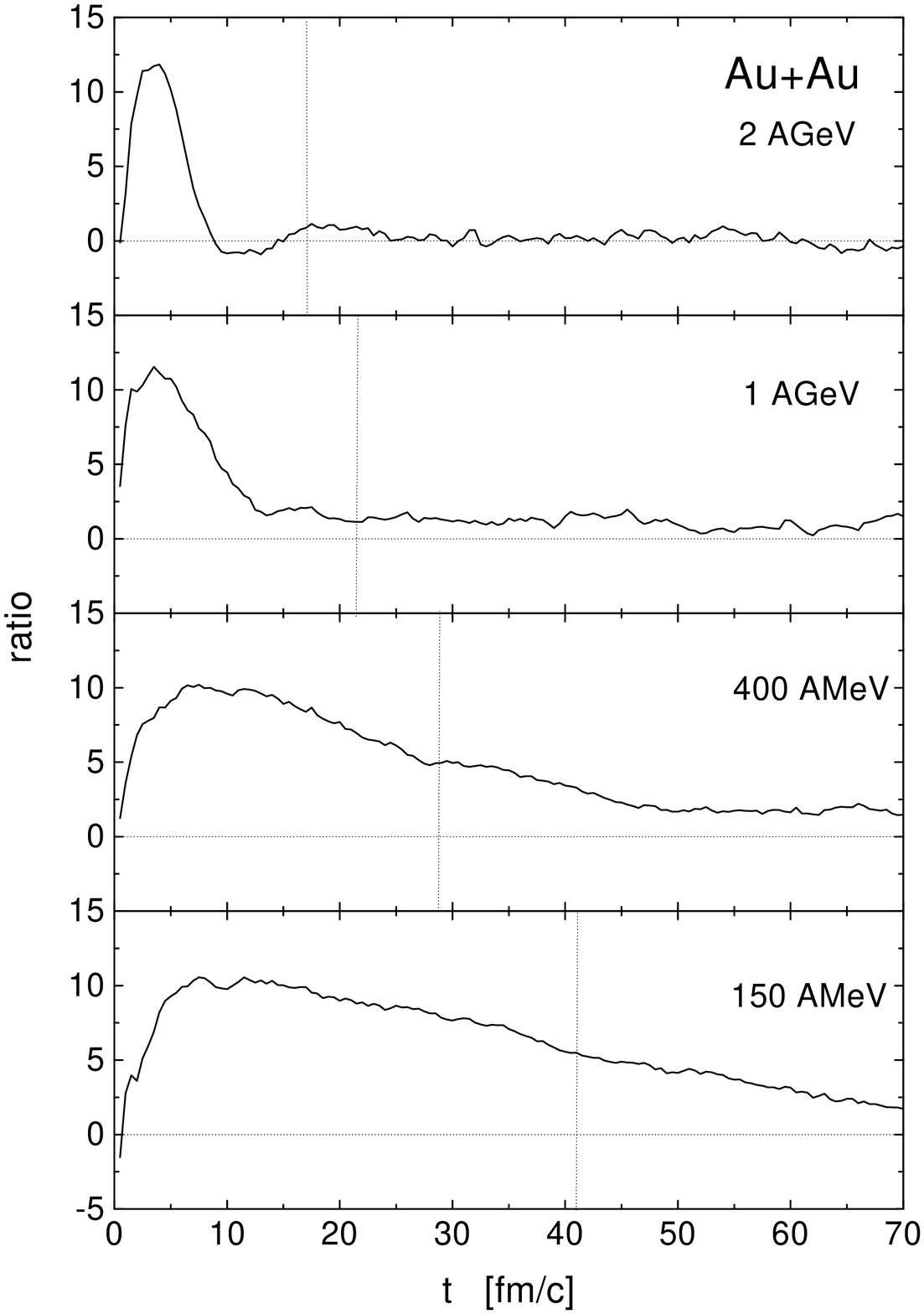}
\includegraphics[angle=90,width=16cm]{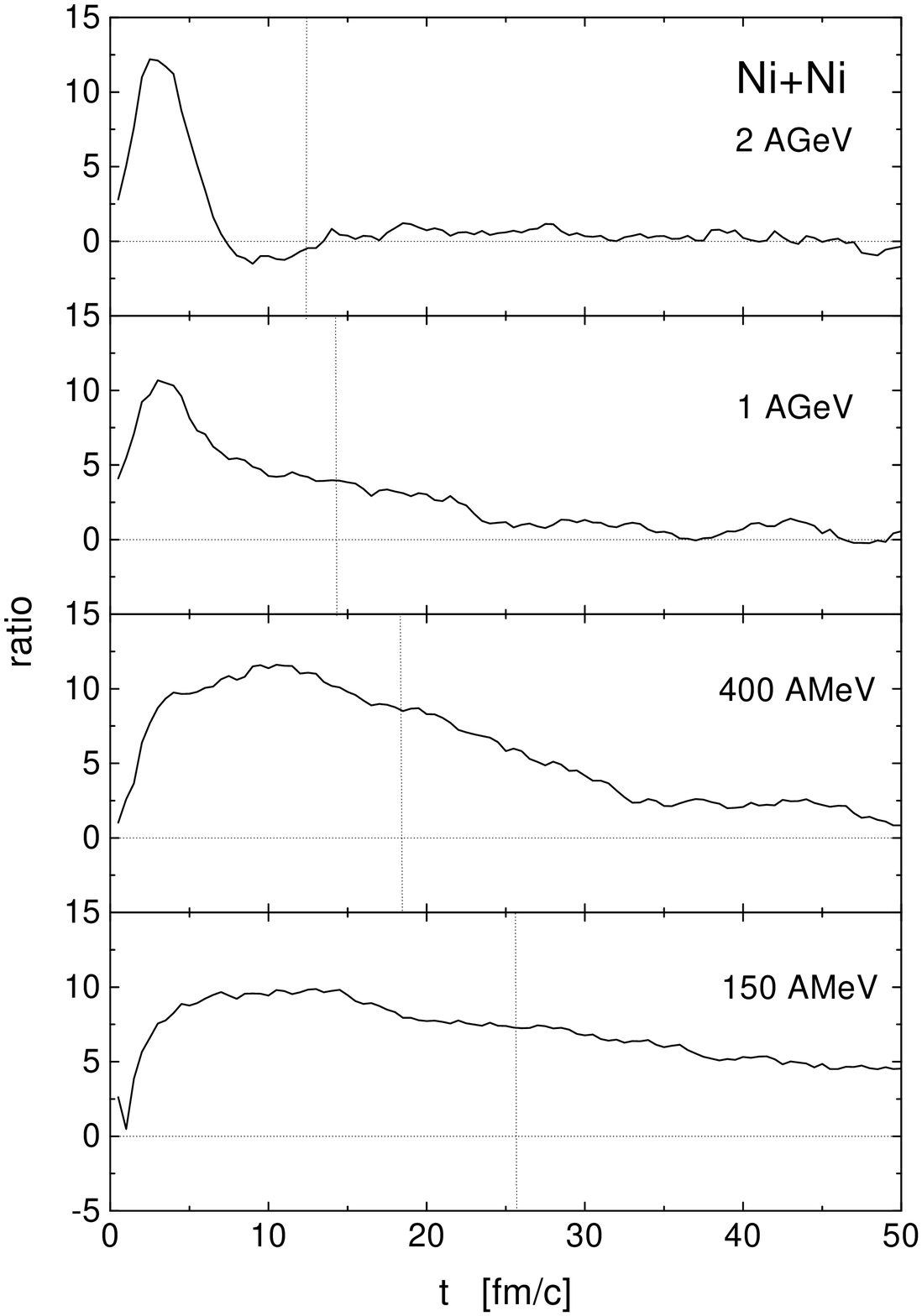}
\vspace{2cm}
\caption{}
\label{eq_auau_nini}
\end{figure}

\begin{figure}[!ht]
\centerline{\includegraphics[angle=90,width=18cm]{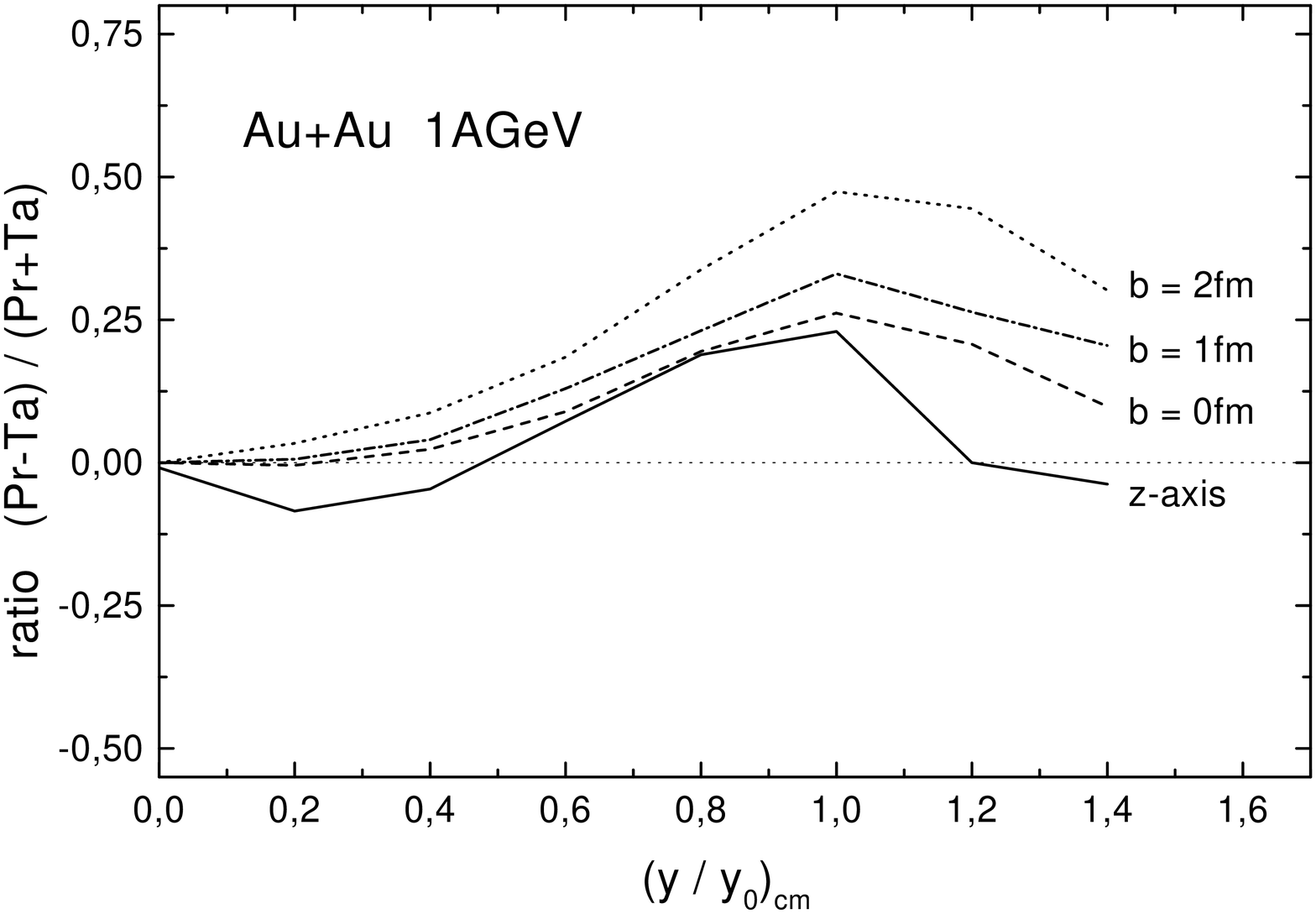}}
\caption{}
\label{mixing_b}
\end{figure}

\begin{figure}[!ht]
\centerline{\includegraphics[angle=90,width=18cm]{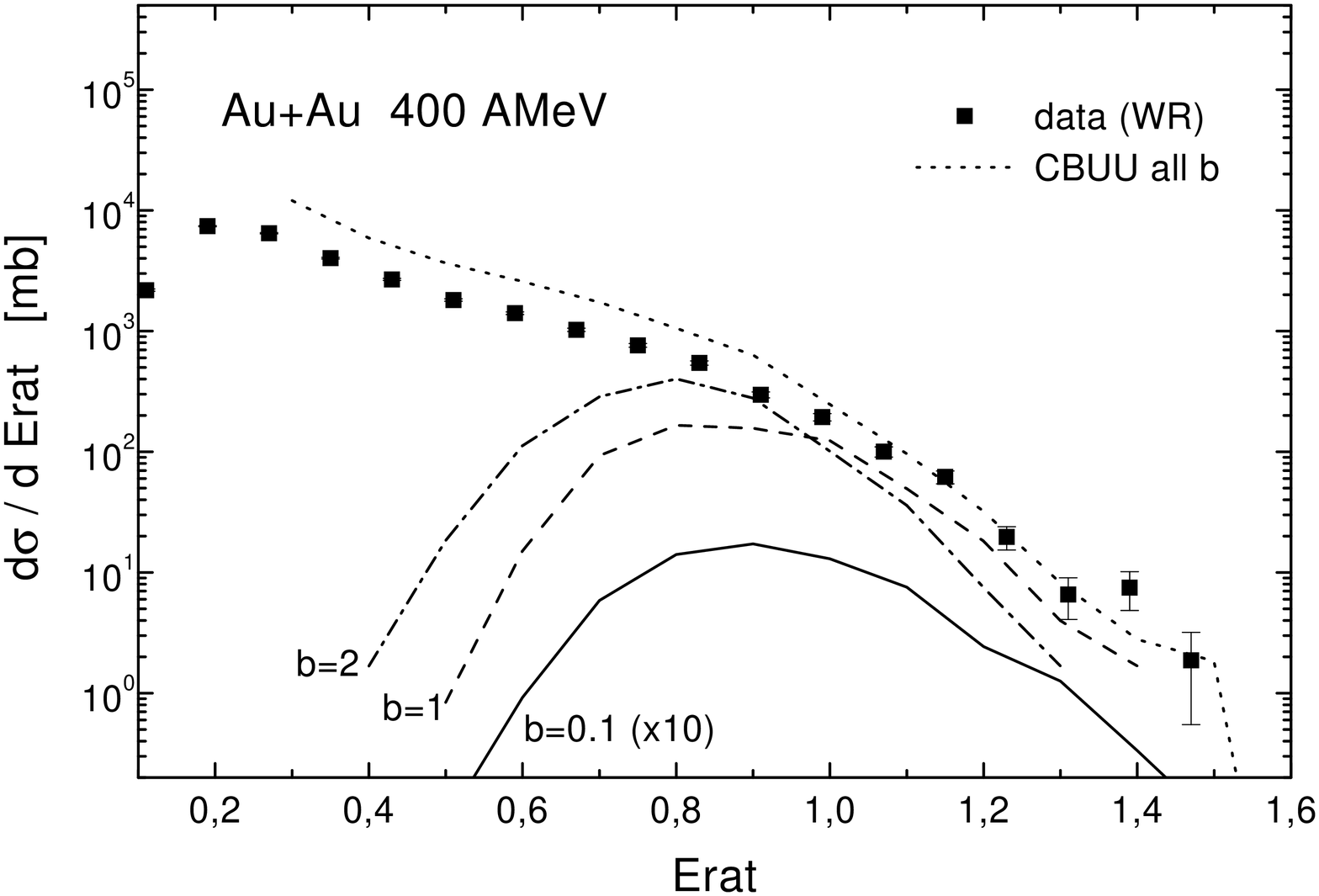}}
\caption{}
\label{erat_b}
\end{figure}

\begin{figure}[!ht]
\centerline{\includegraphics[angle=90,width=18cm]{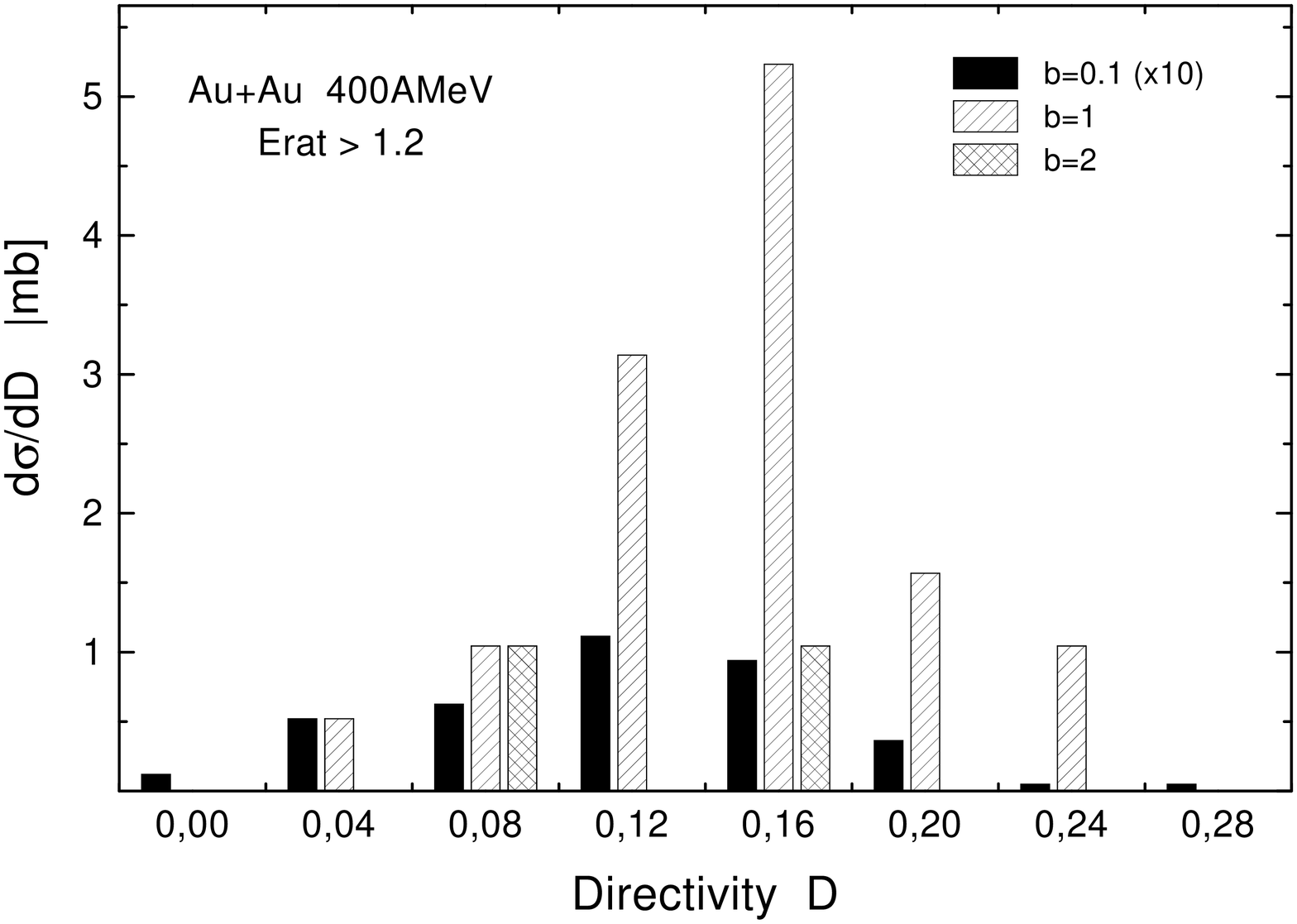}}
\caption{}
\label{direc_erat}
\end{figure}

\begin{figure}[!ht]
\centerline{\includegraphics[angle=90,width=18cm]{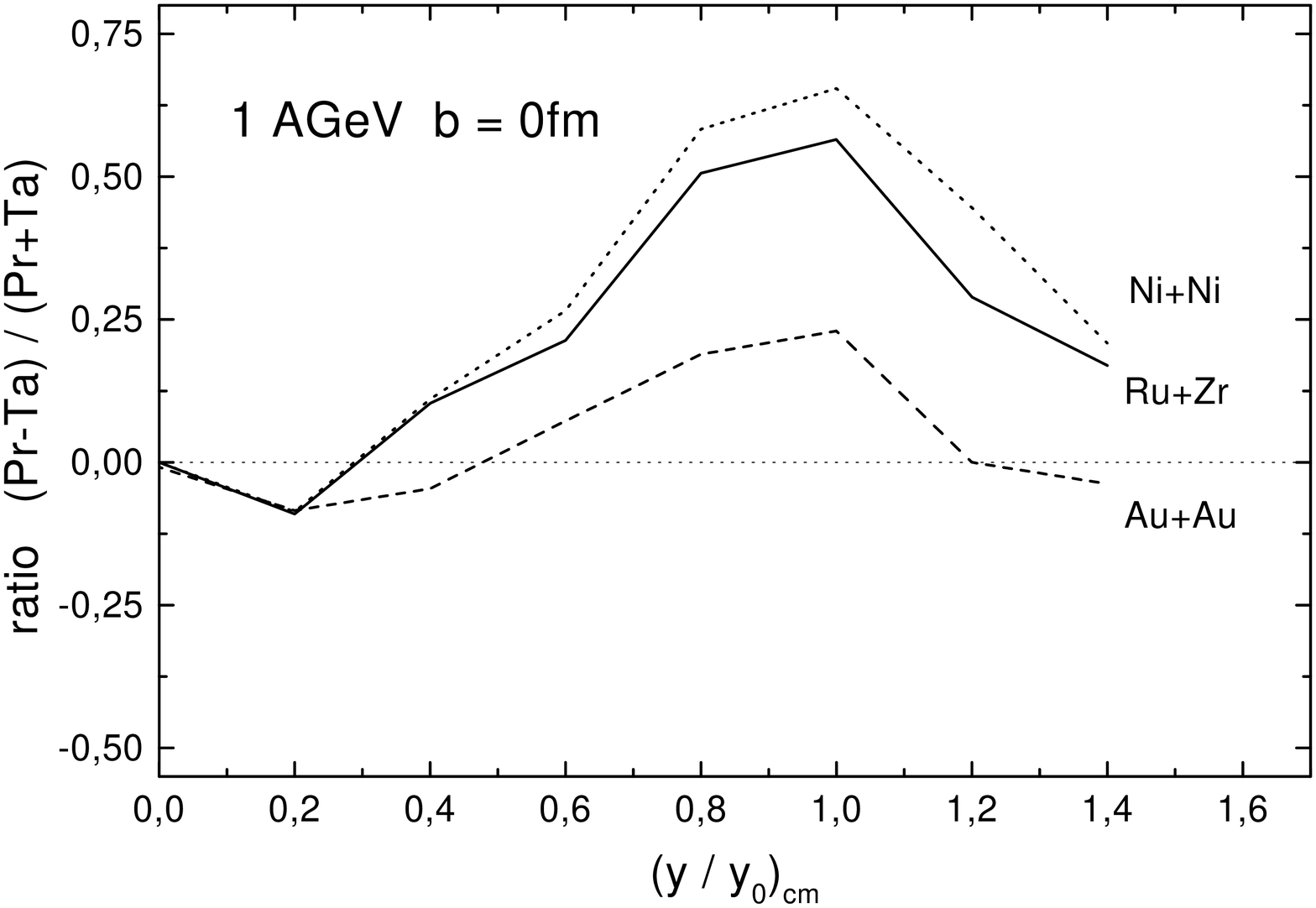}}
\caption{}
\label{transp_m}
\end{figure}

\begin{figure}[!ht]
\centerline{\includegraphics[angle=90,width=18cm]{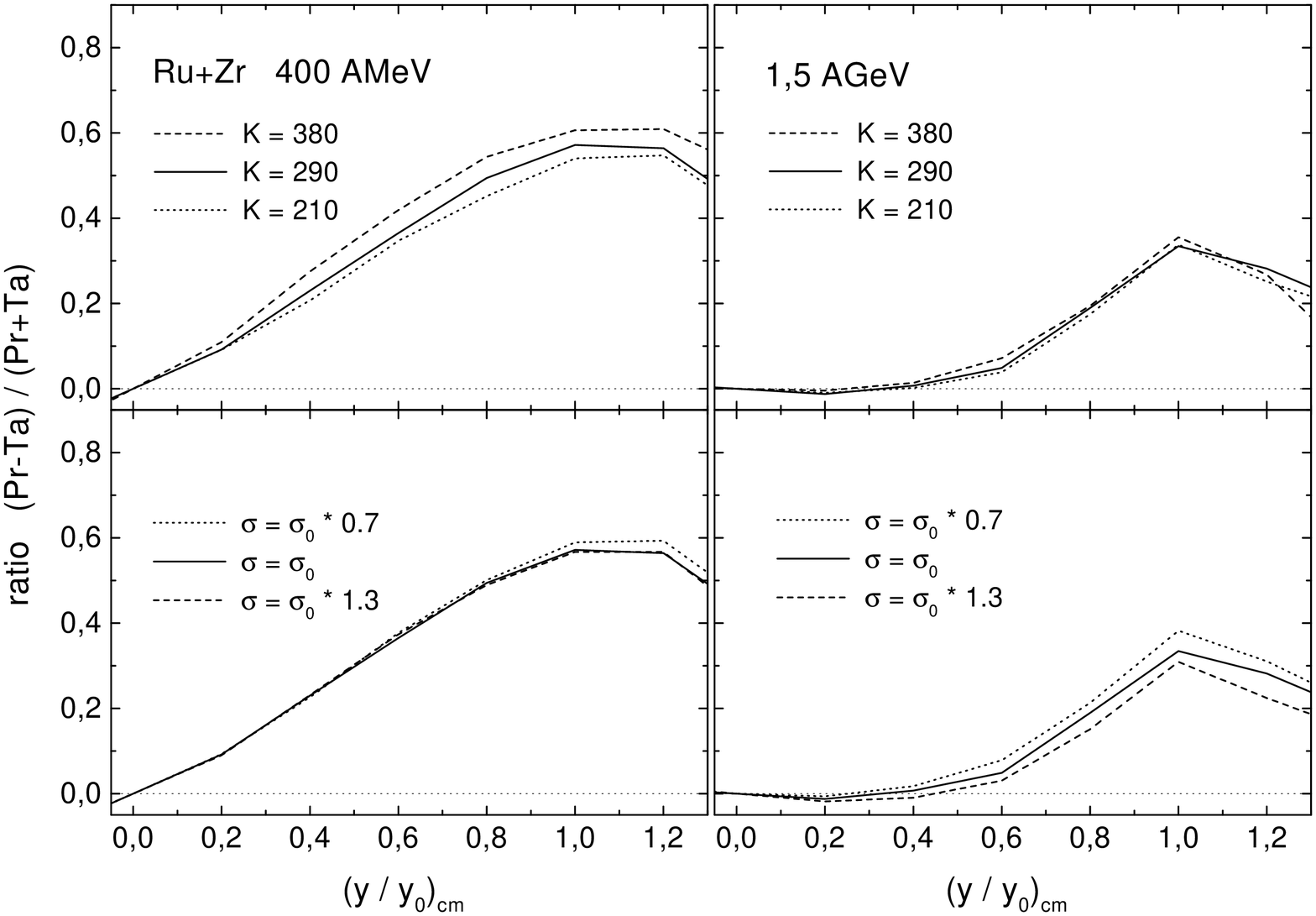}}
\caption{}
\label{ruzr_b0}
\end{figure}

\begin{figure}[!ht]
\centerline{\includegraphics[width=16cm]{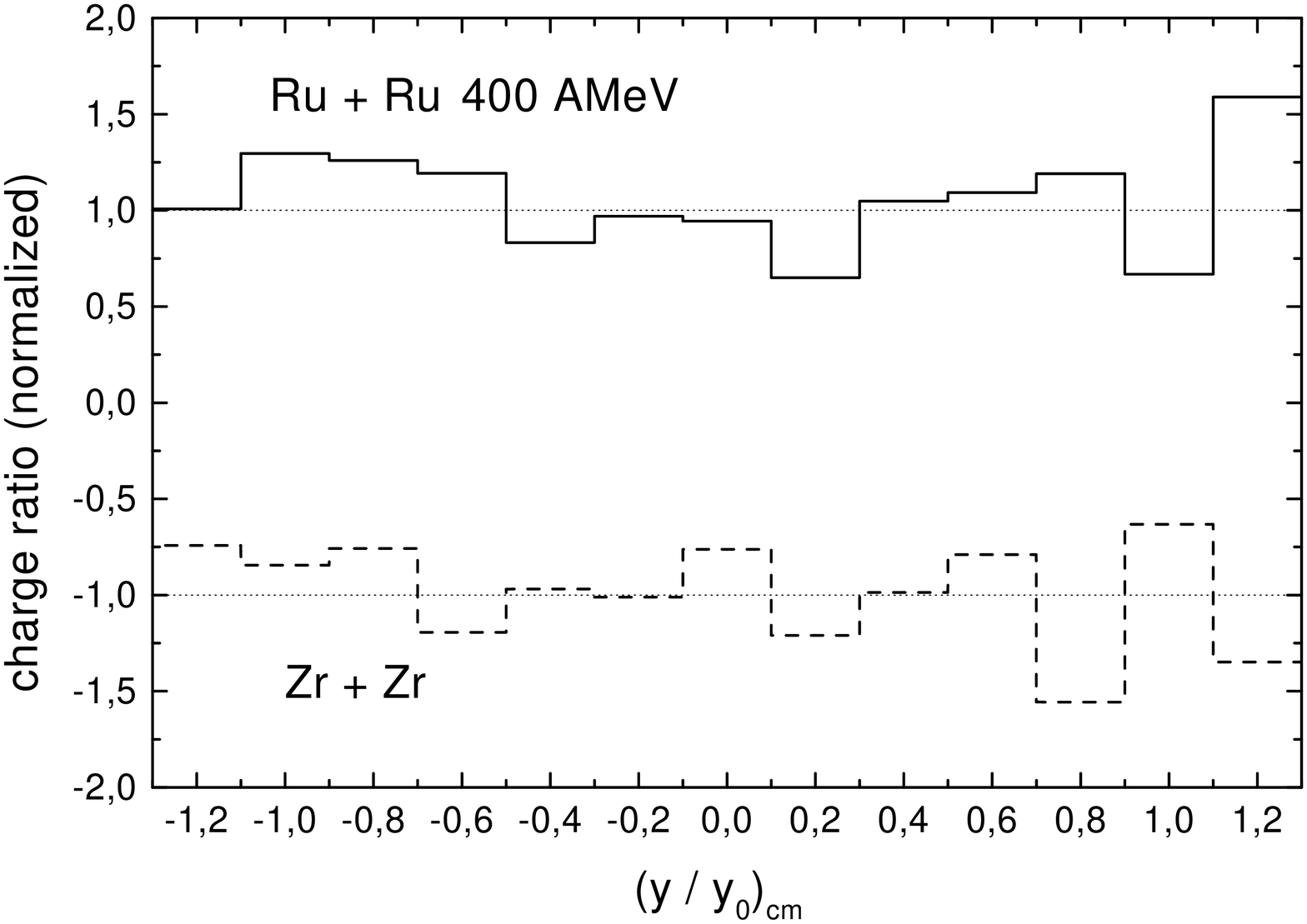}}
\centerline{\includegraphics[width=16cm]{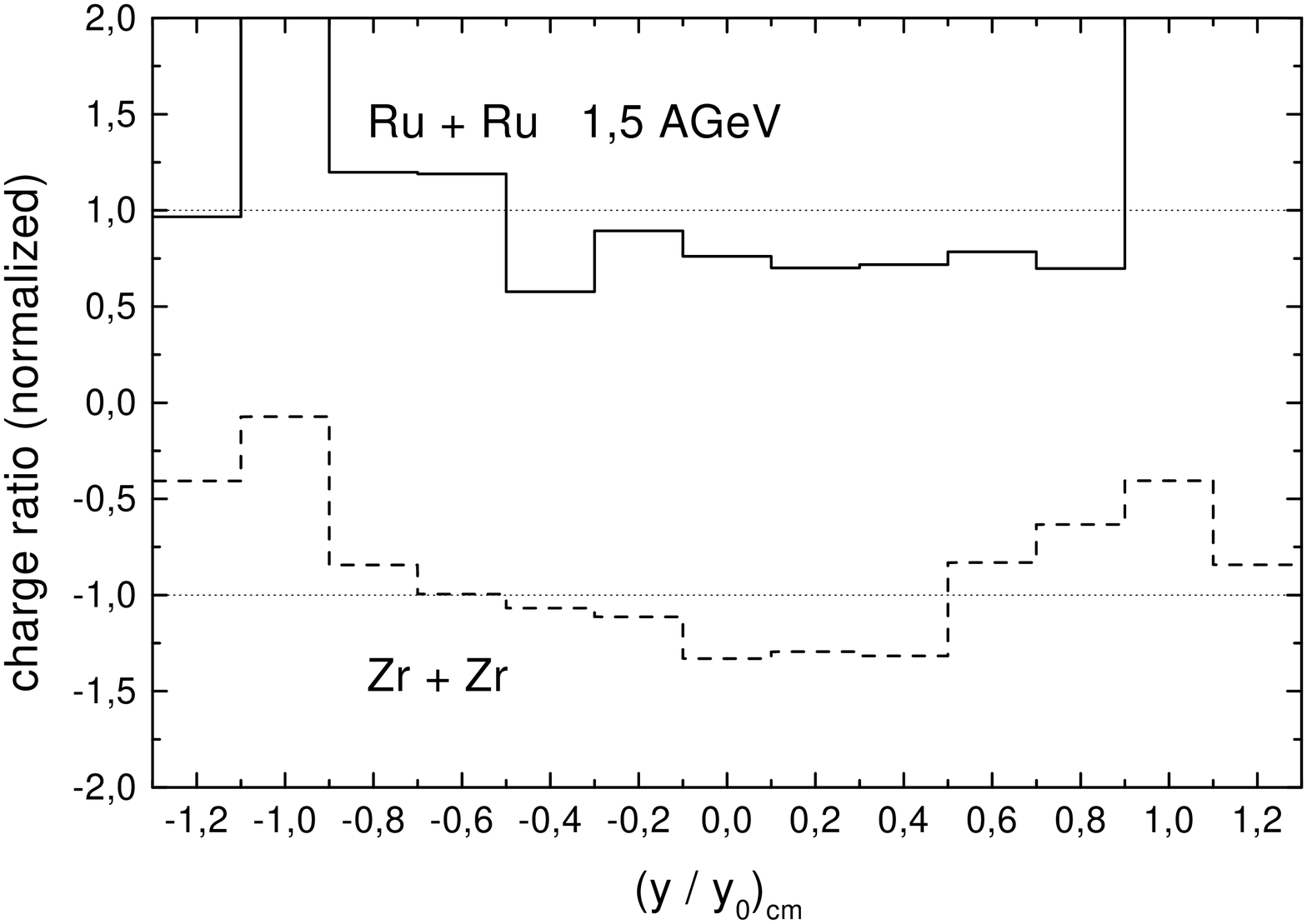}}
\vspace{2cm}
\caption{}
\label{ruru}
\end{figure}

\begin{figure}[!ht]
\centerline{\includegraphics[angle=90,width=18cm]{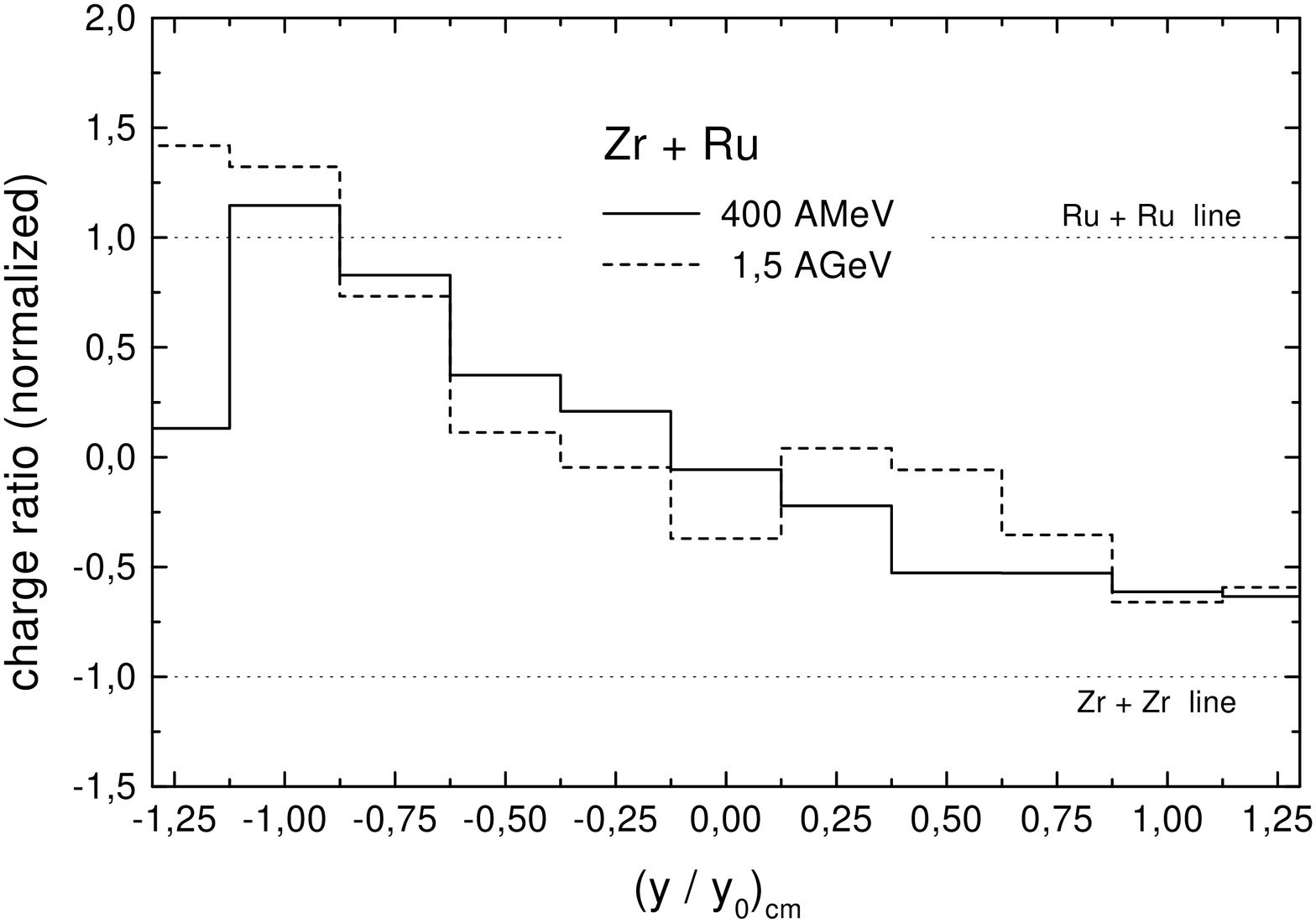}}
\caption{}
\label{ruzr_pn}
\end{figure}

\end{document}